\documentclass[10pt]{IEEEtran}
\usepackage{amsmath}
\usepackage{amsfonts}
\usepackage{amssymb}
\usepackage[ruled,vlined]{algorithm2e}
\usepackage{caption}
\usepackage{multirow}
\usepackage{mathtools}  
\usepackage{amsthm}
\usepackage[]{algorithm2e}
\usepackage{booktabs}
\usepackage{subfigure}
\usepackage{tabulary}
\usepackage{footnote}
\usepackage[export]{adjustbox}
\usepackage{booktabs}
\usepackage[justification=centering]{caption}
\usepackage{float}
\usepackage{comment}

\usepackage{cases}
\usepackage{graphicx}
\usepackage{cite}
\usepackage{multicol}
\usepackage{xcolor}
\usepackage{graphicx}
\graphicspath{{./}{figures/}}
\newcommand\numeq[1]%
  {\stackrel{\scriptscriptstyle(\mkern-1.5mu#1\mkern-1.5mu)}{=}}
\usepackage{stfloats}% <-- added
\usepackage{cuted}
\usepackage{lipsum}  

\usepackage{subfigure}

\title{Revisiting Multi-User Downlink in IEEE 802.11ax: A Designers Guide to MU-MIMO}

\author{\text{Liu Cao}, \text{Lyutianyang Zhang}, \text{Sumit Roy}, \IEEEmembership{Fellow,~IEEE}, \text{Sian Jin}
\thanks{Liu Cao, Lyutianyang Zhang and Sumit Roy are with the Department of Electrical \& Computer Engineering, University of Washington, Seattle, WA, USA (e-mail:\{liucao, lyutiz, sroy\}@uw.edu). Sian Jin is with MathWorks, Natick, MA, USA (e-mail: sianjin@mathworks.com). (\emph{Corresponding author: Lyutianyang Zhang})}}% 

\begin{document}

\maketitle

\begin{abstract}
Downlink (DL) Multi-User (MU) Multiple Input Multiple Output (MU-MIMO) is a key technology that allows multiple concurrent data transmissions from an Access Point (AP) to a selected sub-set of clients for higher network efficiency in IEEE 802.11ax. However, DL MU-MIMO feature is typically turned off as the default setting in AP vendors' products, that is, turning on the DL MU-MIMO may not help increase the network efficiency, which is counter-intuitive. In this article, we provide a sufficiently deep understanding of the interplay between the various underlying factors, i.e., CSI overhead and spatial correlation, which result in negative results when turning on the DL MU-MIMO. Furthermore, we provide a fundamental guideline as a function of operational scenarios to address the fundamental question ``when the DL MU-MIMO should be turned on/off".
\end{abstract}

% Multi-User Multiple Input Multiple Output (MU-MIMO) is a key technology that allows multiple concurrent data transmissions from an Access Point (AP) to a group of users without significant interference or degradation in service quality. While on the Uplink, this implies the use of trigger-based OFDMA, in this work we focus on Downlink Multiuser MIMO (MIMO). Specifically, as the selected subset of clients for MIMO on downlink are closer to each other in dense overlapped networks, increased spatial correlation will lead to significant inter-user and inter-stream interference in DL MU/MIMO channel capacity. Moreover, channel sounding overhead affects both SU and MU aggregate throughput, in particular, a significant increase in spatial correlation with the dimensionality of MIMO. In this work, we show how spatial correlation and channel sounding overhead relate to SU/MU throughput following IEEE 802.11ax channel sounding. We also provide a fundamental operation guideline for DL SU/MU-MIMO.

\section{Introduction} 
\label{introduction}

%While on the Uplink, this implies use of trigger-based OFDMA, in this work we focus on Downlink Multi-User Multiple Input Multiple Output (MIMO). Legacy single user MIMO (SU-MIMO) - the precursor to MU-MIMO - laid the groundwork by allowing the transmission of multiple data streams from an AP equipped with multiple antennas to a single client device on downlink. 

IEEE 802.11ax (Wi-Fi 6) marked a significant evolution milestone via the introduction of Multi-User (MU) communication modes (in contrast with legacy Single-User (SU)  communication) for both Uplink (UL) and Downlink (DL) in tri-band (2.4/5/6 GHz) \cite{11ax}. For the Uplink, this implies the use of trigger-based OFDMA; in this article, we focus solely on DL Multi-User (MU) Multiple Input Multiple Output (MU-MIMO).  Legacy Single-User MIMO (SU-MIMO) - the precursor to MU-MIMO - laid the groundwork by allowing transmission of multiple spatial streams from an access point (AP) equipped with multiple antennas to a single client device on downlink.  With the proliferation of wireless client devices, a single Wi-Fi network access point (AP) can have multiple associated stations (STAs) \cite{cisco,qualcom}. With multi-antenna clients \footnote{However, the number of antennas at the AP always exceeds the number of antennas at a client.}, it is feasible via DL Transmit Beamforming (TxBF) at the AP to send multiple streams to multiple STAs simultaneously (DL MU-MIMO). 

\begin{figure}[tp]
    \centering
\includegraphics[width=.48\textwidth]{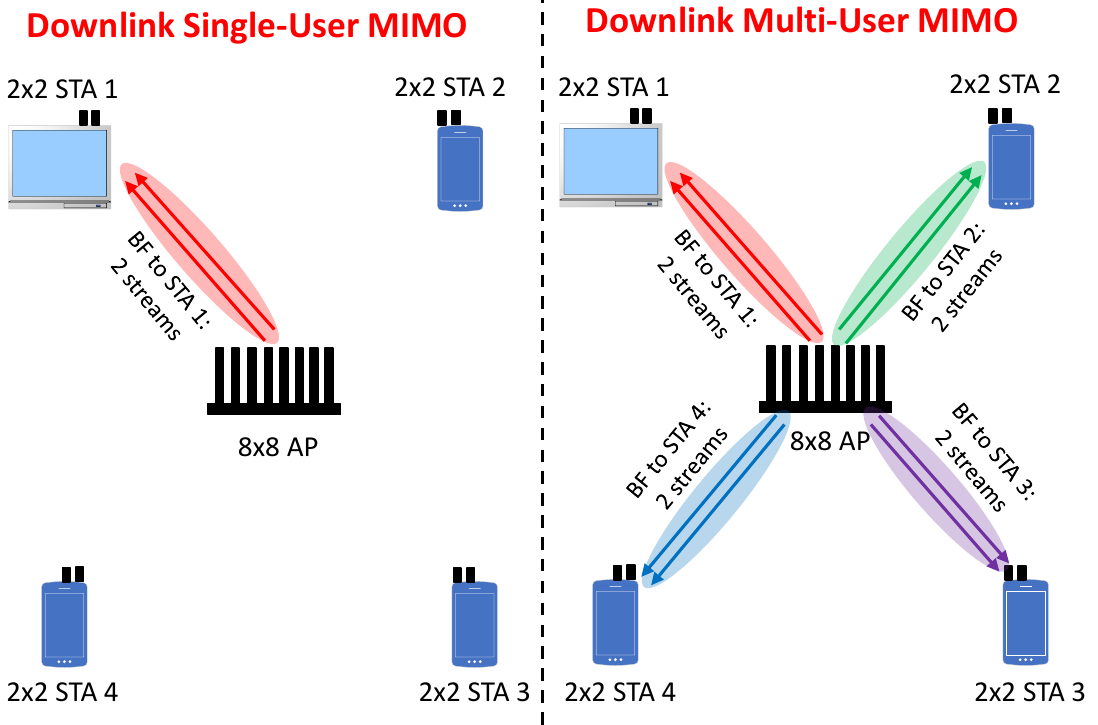}
    \caption{SU-MIMO vs MU-MIMO on Downlink Operations.}
    \label{fig:sys}
\end{figure}

A typical configuration \cite{white_signal, white_arista} such as Fig.~\ref{fig:sys} assumes an 8 x 8 AP (e.g., NetGear RAXE500) and 2 x 2 STAs (e.g., iPhone 15 and MacBook Air), implying that a single downlink transmission opportunity can potentially send a total of $8$  spatial streams \footnote{Note that Wi-Fi 5 (IEEE 802.11ac) included support for MU-MIMO but limited to 4 streams on only 5 GHz downlink operation; whereas Wi-Fi 6 supports up to 8-stream on 2.4/5/6 GHz uplink/downlink operations.} to a selected sub-set of clients, e.g. each 2 streams to each selected four STAs. While DL SU-MIMO results in scaling
of per-user throughput as a result of multi-stream transmission, its benefits are limited by the fact that most clients support either 1 or 2 spatial streams (i.e., a total of 2-stream transmissions in DL SU-MIMO in Fig.~\ref{fig:sys}). By contrast, it is evident that in dense overlapped network scenarios - such as the enterprise or residential cluster - DL MU-MIMO provides a natural pathway to increasing network efficiency (aggregate network throughput) by {\em enabling simultaneous transmissions of multiple streams to multiple clients} (i.e., a total of 8-stream transmissions in DL MU-MIMO in Fig.~\ref{fig:sys}), with appropriate choice of the user sub-set and TxBF to minimize inter-user/inter-stream interference. 

% This capability leads to two significant benefits: first, the achievement of higher data rates when communicating with a single client with more spatial streams, and second, supporting simultaneous communication with multiple STAs.

Despite the promise of MU-MIMO for improved network capacity via simultaneous transmission to multiple users on downlink\footnote{There exists an analogous feature for the uplink: trigger-based OFDMA whereby a $20$ MHz channel may be shared synchronously by multiple users. However, consideration of UL OFDMA is beyond the scope of this article.}, real-world user testing has revealed significant challenges. A noticeable discrepancy exists between the theoretical speeds advertised by manufacturers who incorporate DL MU-MIMO and the actual throughput measured in specific conditions \cite{cambium,choi2019use, Extreme}. An industry test report \cite{higgins} showed that turning on MU-MIMO resulted in 58\% aggregate throughput {\em loss} compared to SU-MIMO when pairing 4 x 4 Broadcom-based router with 2 x 2 Qualcomm-based STAs. An earlier research study \cite{sur2016practical} demonstrated that DL SU-MIMO achieves 16.8\% to 42\% higher aggregated throughput MU-MIMO based on a test of a commercial 4 x 4 MU-MIMO-capable 802.11ac 5 GHz radio with 1 x 1 Xiaomi Mi 4i smartphones. Such variation in results is attributable to various factors at play, including the complex interplay of channel state information (CSI) overhead, device capabilities and environmental (propagation) conditions as a function of user location. In this article, we chose the IEEE 802.11ax indoor channel model \cite{liu2014ieee}, widely used by the industry and academia, for a foundational exploration of  DL SU/MU-MIMO throughput. Specifically, as the selected sub-set of clients for MU-MIMO on downlink are closer to each other in dense networks, increased spatial correlation will lead to significant inter-user and inter-stream interference in DL MU-MIMO. Thus overall network throughput degrades unless counteracted by a combination of inter-user interference cancellation and user selection algorithms \cite{1603708, bjornson2014optimal}. Moreover, CSI overhead affects both SU and MU aggregate throughput; in particular, CSI overhead increases significantly with the dimensionality of MU-MIMO. In turn, this implies that any MU-MIMO design must carefully consider the issue of (optimal) channel sounding periodicity when confronted with channel time variations\footnote{Further consideration of this topic is beyond the scope of this article.}.

The lack of a sufficiently deep understanding of the interplay between the various underlying factors discussed has resulted in AP vendors turning off the DL MU-MIMO feature as default setting in their products, reflecting the current ambivalence surrounding DL MU-MIMO. The primary purpose of this article is therefore to provide {\em new insights underlying the fundamental question}: ``\textbf{when should DL MU-MIMO be turned on/off}" as a function of the operational scenario. By a combination of analysis and computation/simulation, we attempt to answer the above question by  
\begin{itemize}
    \item Identifying set of conditions where DL SU-MIMO outperforms MU-MIMO and vice-versa;
    \item Provide broad `rules of thumb' regarding use of DL MU-MIMO in current/future Wi-Fi systems. 
\end{itemize}

The rest of this article is organized as follows. Section \ref{secII} introduces the impact of DL SU and MU CSI overhead differences on their effective channel capacity; In Section \ref{secIII}, we explore the impact of spatial correlation on the MU channel capacity under the IEEE 802.11ax indoor channel model. In Section \ref{secIV}, a design guideline table for DL MU-MIMO is proposed by unifying the factors discussed in Section \ref{secII} and \ref{secIII}.  Finally, Section \ref{secV} concludes this article.

\section{Factor 1: CSI Overhead}
\label{secII}
 In 802.11ax DL transmission, AP is the transmitter which is called the beamformer, while a STA is the receiver which is called the beamformee. Beamforming depends on channel calibration procedures, called channel sounding in the 802.11ax standard. The channel sounding allows the beamformer to gather the beamforming report(s) that characterize the beamformee location(s) and to transmit the streams toward the precise direction of the beamformee(s).

\begin{figure*}[t]
    \centering
\includegraphics[width=.85\textwidth]{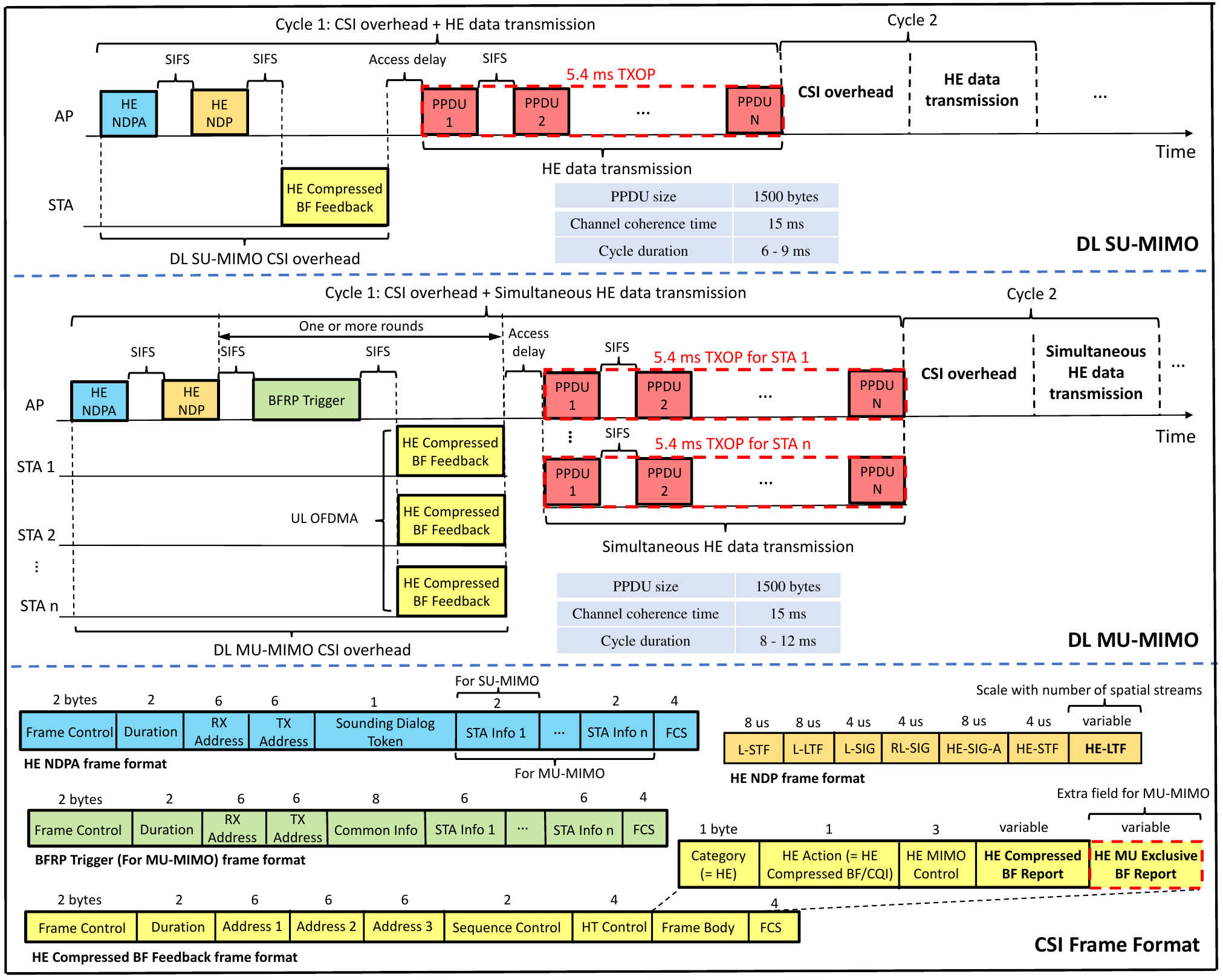}
    \caption{IEEE 802.11ax Channel Sounding followed by High-Efficiency (HE) Data Transmission.}
    \label{fig:he}
\end{figure*}

\subsection{DL SU/MU-MIMO channel sounding}
DL SU-MIMO is indicated by codebook info 0 in the High-Efficiency (HE) MIMO control field. As Fig. \ref{fig:he} shows, its channel sounding process consists of four major steps: 
\begin{itemize}
    \item The beamformer begins the process by broadcasting a Null Data Packet Announcement (NDPA) frame, which is used to gain control of the channel and identify the intended beamformee. 
    \item The beamformer next transmits a Null Data Packet (NDP) to beamformee after a Short Interframe Space (SIFS). NDP is an empty frame that only contains the Physical Layer Protocol Data Unit (PPDU) header. The received NDP is used for channel estimation by analyzing the OFDM training symbols, called HE-LTF, whose length is a variable that depends on the number of spatial streams.
    \item Following receipt of the NDP, the beamformee responds with a BF feedback matrix in a compressed form. The BF feedback matrix instructs how the beamformer should steer the data frame to the beamformee with higher energy. Codebook information in the HE MIMO Control field provides the resolution schemes for compressing the BF feedback matrix.  
    \item The beamformer receives and recovers the compressed feedback matrix that is further used as the steering matrix to direct HE data transmissions toward the beamformee.
\end{itemize}

By contrast, DL MU-MIMO, indicated by codebook info 1 in the HE MIMO control field, follows the similar channel sounding protocols as the SU-MIMO, however, several \textbf{major} differences exist:
\begin{itemize}
    \item NDPA frame format: A HE NDPA frame in MU-MIMO includes multiple STA Info fields, one for each beamformee, while the NDPA frame in SU-MIMO only carries a single STA Info field.
    \item BF Report Poll (BFRP) trigger frame: The compressed BF feedback in SU-MIMO comes right after the NDP. However, the beamformer in DL MU-MIMO must use a control frame - BFRP Trigger frame that instructs each beamformees to transmit the BF feedback simultaneously. The AP may transmit other BFRP Trigger frames to gather more feedbacks if necessary.  
    \item Compressed BF feedback frame format: The HE MU Exclusive BF report is an extra field at the end of the frame for MU-MIMO, which thereby introduce extra CSI overhead; 
    \item BF Feedback transmission: The BF feedback in SU-MIMO is transmitted over the UL OFDM while they are transmitted over the UL OFDMA in MU-MIMO.  
\end{itemize}

% In addition, channel sounding should be done periodically to provide the AP with accurate channel measurements. Experience with SU-MIMO shows that a channel measurement is good for approximately 100 ms \cite{gast2013802}. While a typical channel sounding in MU-MIMO varies between 40 ms and 10 ms \cite{gast2013802}. It may be even less than 10 ms \cite{choi2016sounding}.

\subsection{CSI Overhead Comparison}

CSI overhead in DL SU/MU-MIMO can be calculated based on the CSI frame format indicating each sub-field size, as shown in Fig. \ref{fig:he}. In particular, CSI overhead is dominated by HE compressed BF feedback that contains the {\em HE compressed BF report} (as well as the extra sub-field - {\em HE MU Exclusive BF report} in MU-MIMO). The compressed BF report contains the compressed CSI for each sub-carrier, i.e., the \textbf{V-matrix} or steering/precoding matrix\footnote{The Null-steering step based on zero-forcing (ZF) and minimum mean square error (MMSE) approaches \cite{1603708,rappaport2024wireless}, used for precoding in DL MU-MIMO are not implemented in real AP products \cite{sur2016practical,choi2019use} because those can be implemented only if the full CSI is obtained, whereas the feedback V-matrix provides only partial CSI. Besides, null-steering step incurs additional computational complexity and thus chipset cost for the AP.} used for digital beamforming. V-matrix is obtained by a) applying the singular value decomposition (SVD) to the full CSI, and b) compressing it to specific {\em Givens rotation angles} to reduce the amount of required bits.  The compressed size of the V-matrix depends on the total number of Givens rotation angles as well as the number of bits used to quantize each angle, as defined in IEEE 802.11ax specification. In general, the larger the V-matrix dimension, the more the number of angles. Meanwhile, the number of bits to quantize each angle is indicated by the sub-field of codebook information choice with 1 bit in the {\em HE MIMO Control field}. Thus both SU- and MU- have two codebook information choices \cite{11ax}, however, MU-MIMO uses more bits than SU-MIMO to quantize a single angle by using the same codebook information. For instance, if codebook information bit is set to 0, the number of bits to quantize an angle in SU-MIMO is 4 or 2 while they are 7 or 5 in MU-MIMO \cite{11ax}, implying that the compressed V-matrix in MU-MIMO has larger overhead compared to SU-MIMO. In addition, the HE compressed BF report size also scales with the number of spatial streams and the number of sub-carriers. The  MU-Exclusive BF report in MU-MIMO contains the delta SNR per sub-carrier, which represents the difference from the average SNR. The MU-Exclusive BF report represents the spatial characteristics for each sub-carrier caused by the environment, the size of which scales with the number of subcarriers. Since the 802.11ax specification does not detail how this information is exploited in the design of the beamformer, its implementation is chip vendor dependent.

\begin{comment}
\begin{figure*}[htp]
    \centering
\includegraphics[width=.85\textwidth]{Table.pdf}
    \caption{The variables that dominate the CSI overhead in the HE compressed BF feedback.}
    \label{fig:table}
\end{figure*}
\end{comment}
    %     \begin{itemize}
    %     \item If the Feedback Type subfield indicates SU: Set to 0 to indicate $b_{\phi} = 4$ and $b_{\psi} = 2$; or set 1 to indicate $b_{\phi} = 6$ and $b_{\psi} = 4$;   
    %     \item If the Feedback Type subfield indicates MU: Set to 0 to indicate $b_{\phi} = 7$ and $b_{\psi} = 5$; or set 1 to indicate $b_{\phi} = 9$ and $b_{\psi} = 7$;   
    % \end{itemize}

% CSI overhead scales with
% \begin{itemize}
%     \item \textcolor{red}{\# of STAs + \# of antennas}
%     \item \textcolor{red}{channel sounding frequency}
% \end{itemize}
As discussed, channel sounding procedures introduce a significant cost in airtime because the sounding exchange must be completed before a beamformed data transmission can occur.  Therefore, if the MU-MIMO BF gain is not sufficient to offset the airtime consumed by the sounding exchange, MU-MIMO throughput can be lower than the SU-MIMO in some operational scenarios. 

\begin{figure}[t]
    \centering
\includegraphics[width=.48\textwidth]{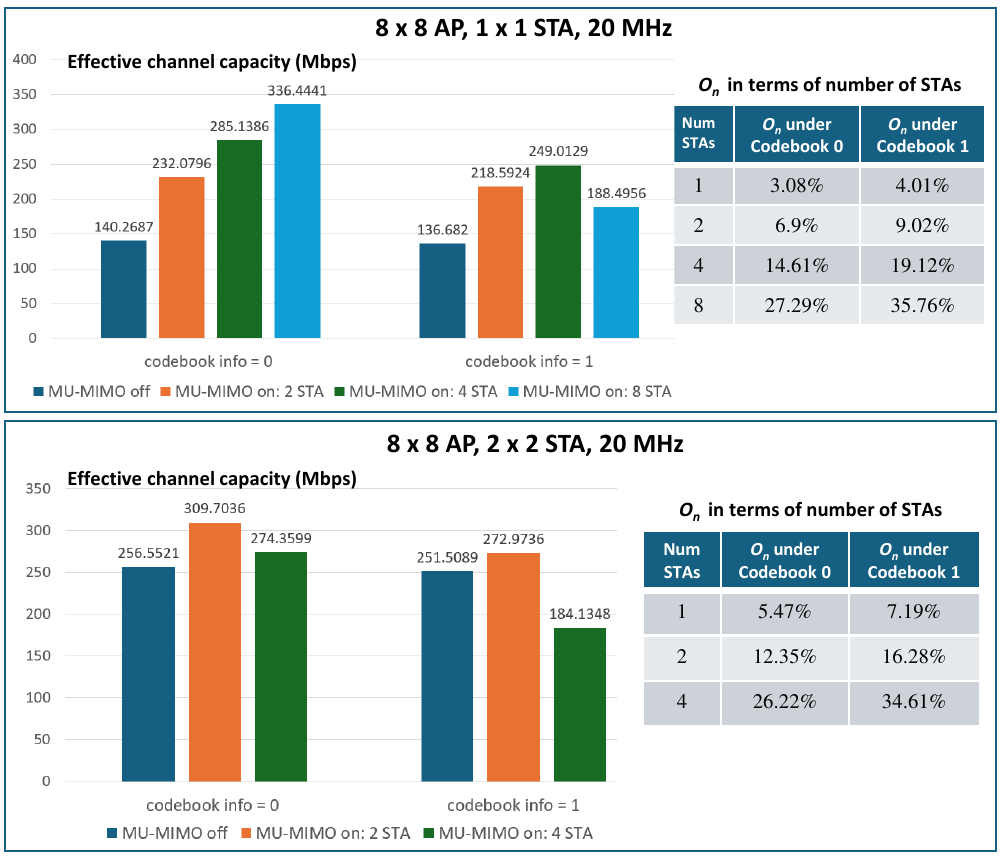}
    \caption{Effective Channel Capacity impacted by CSI Overhead. Average 25 dB SNR at the single STA in SU-MIMO. } 
    \label{fig:ohExample}
\end{figure}

As Fig. \ref{fig:he} shows, a cycle of CSI overhead and HE data transmission is repeated in both DL SU and MU-MIMO. In each cycle, the transmitted data for each STA is filled in one Transmit opportunity (TXOP) comprised of multiple back-to-back PPDUs (e.g., 1500 bytes) in SIFS burst mode. Thus the data transmission duration is the maximum TXOP limit ($5.4$ ms) compared to which the duration of access delay is negligible (typically less than a few hundred microseconds), as long as the number of STAs is not excessively large. If we assume STA's walking speed equal to 2 mph, the resulting channel coherence time \footnote{Channel coherence time is defined as the time duration over which the channel is considered to be not varying.} (15 ms) will be greater than any one cycle duration. Hence, it is reasonable to assume a block fading channel for each cycle, i.e., channel capacity is fixed in a cycle while varying across different cycles. We will use the {\em effective channel capacity} $C_{\text{eff}}$  to compare the SU and MU- performance as a function of CSI overhead - defined as the average channel capacity over both CSI overhead (zero channel capacity) duration and HE data transmission duration (non-zero channel capacity), given by 
\begin{equation}
    C_{\text{eff}} = \frac{1}{N}  \sum_{n,k} (1 - O_{n}) \cdot C_{n,k}, \label{eq:C}
\end{equation}
where $N$ denotes the total number of cycles, $C_{n,k}$ denotes the Shannon channel capacity \cite{rappaport2024wireless} of the $k$-th STA in the $n$-th cycle, assumed to be constant due to the block fading channel. $O_{n}$ is the ratio of CSI overhead airtime to the cycle duration for the $n$-th cycle. Eq. (\ref{eq:C}) applies to DL SU-MIMO when the size of $k$ is 1. Note that $C_{n,k}$  varies across $n$ due to time-varying channels\footnote{For the pure analysis of CSI overhead in this section, the inter-user interference determined by spatial correlation is assumed to be zero. Thus $C_{n,k}$ in terms of $n$  changes only due to the channel gain variations rather than variations in inter-user interference. Then, $C_{\text{eff}}$ shown in Fig. \ref{fig:ohExample} is the maximum effective channel capacity that DL MU-MIMO can reach.}, but $O_{n}$ is independent of $n$ in our model since we assume a specific setup (i.e., MIMO dimension, codebook information, and the number of selected STAs, TXOP duration).

% \begin{equation}
%     \text{C}_{\text{eff}} = \frac{1}{N}  \sum_{n,k,i} (1 - O_{n}) \cdot B \cdot\text{log}_2\bigg(1+\text{SNR}_{n, k,i}\bigg), \label{eq:C}
% \end{equation}

% \begin{equation}
%     O_{n} = \frac{\text{Duration of CSI overhead}}{\text{Duration of CSI overhead} + \text{Duration of Data Transmission}}
% \end{equation}

% (23 PPDUs) with MCS 5, 20 MHz bandwidth and 0.8 $us$ Guard Interval (GI)

% SNR from 25dB to 20dB with sumimo 100ms

% SNR from 20dB to 15dB with mumimo 40ms

% mumimo susceptbile to time varying factor.

% jakes model - IEEE time varying ....

% PPT every point

% In the SU-MIMO case, 1 STA with 1 antenna is connected to this AP over the 20 MHz bandwidth (i.e., $ N_s =$ 242 subcarriers),  In MU-MIMO case N STAs (each with 1 antenna) are connected to the AP over the 20 MHz bandwidth, Suppose SNR = 8 in SU-MIMO

Assuming an 8 x 8 AP, 1 x 1 STA(s), and 20 MHz channel bandwidth as in Fig. \ref{fig:ohExample}, the effective channel capacity does {\em not} grow linearly with the number of STAs. In particular, the effective channel capacity under codebook info 1 is greatly reduced when the number of STAs reaches 8. This can be explained by following reasons:
\begin{itemize}
    \item CSI overhead proportion $O_n$  shown in Fig. \ref{fig:ohExample} grows exponentially with the number of STAs. This is because, on the one hand, an extra field - HE MU Exclusive BF report as a function of the number of sub-carriers and spatial streams - is included in HE compressed BF feedback, incurring extra CSI overhead; On the other hand, the bandwidth is shared using UL OFDMA leads to the lower UL data rate for HE compressed BF feedback transmission per STA; Thus, the DL MU-MIMO CSI overhead becomes significantly higher than SU-MIMO for a large number of STAs;
    \item AP transmit power is divided equally for each STA in DL MU-MIMO. As a result, $C_{n,k}$ in Eq. (\ref{eq:C}) will drop with increasing number of STAs due to the lower transmit power per STA.
\end{itemize}

The same phenomenon repeats for the 8 x 8 AP, 2 x 2 STA, and 20 MHz cases in Fig. \ref{fig:ohExample}; the effective channel capacity is reduced when the number of STAs reaches 4. However, this does not indicate that AP shall not support more STAs due to the lower effective channel capacity. However, inspite of this result, AP vendors may choose to support greater number of STAs on simultaneous DL as that may be independently desirable\cite{cisco}.  It is noteworthy that codebook info 1 (i.e., using more bits to quantize the V-matrix) always has lower effective channel capacity performance than codebook info 0 in Fig \ref{fig:ohExample}. This is because we assume the perfect channel estimation which does not produce channel estimation error under both Codebook info 0 and 1. Thus codebook info 1 with larger CSI overhead always suffers more than codebook info 0.

% It is also noteworthy that the channel sounding period should be more frequent for MU than SU; however, the MU channel sounding also consumes more airtime for CBFR, which necessitates the next-gen optimized channel sounding to strike the balance between MU channel sounding frequency and efficiency.
 % The channel sounding procedure starts with a Null Data Packet (NDP) Announcement (NDPA) followed by an NDP and a Beamforming Report Poll (BRP) frame transmitted by the AP. The different beamformees identified in the BRP reply simultaneously and send HE Compressed Beamforming feedback using UL Orthogonal Frequency-Division Multiple Access (OFDMA). The AP may transmit other BRP Trigger frames to gather more feedback if necessary. An example of the HE channel sounding followed by simultaneous transmission is depicted in Fig. \ref{fig:mumimo_mechnism}. 

% NDP is an empty frame that only contains the PPDU header. 

% To make it work, the access point queries the client device using a sounding protocol to obtain sub-channel information. Sub-channel information can also be sent by the client in unsolicited management frames. Using this information, the access point does three things:

% \label{sec:sys_arc}

% \begin{figure}[ht]
% \centering
%  \subfigure[DL SU-MIMO.]
% {\includegraphics[width=0.48\textwidth]{SUMIMO.pdf}}
% \label{fig:1a}
% \centering \\
% \subfigure[DL MU-MIMO.]{
% \includegraphics[width=0.48\textwidth]{MUMIMO.pdf}}
% \label{fig:1b}
% \caption{802.11ax CSI Exchange followed by Data Transmission.}
% \label{fig:mumimo_mechnism}
% \end{figure}

\section{Factor 2: Spatial correlation}
\label{secIII}

In this section, we investigate the impact of spatial correlation on the SU and MU performance in practical environmental conditions. The spatial correlation among user's is characterized by two key factors: user separation, and distance between AP and STAs. We use the Shannon channel capacity (without CSI overhead) as the metric to investigate the SU and MU throughput as a function of spatial correlation next.
%Unifying the impact of CSI overhead as well as spatial correlation will be discussed in Section \ref{secIV}.

\subsection{Clustered-based multi-path channel model}
\begin{figure}[t]
    \centering
\includegraphics[width=.42\textwidth]{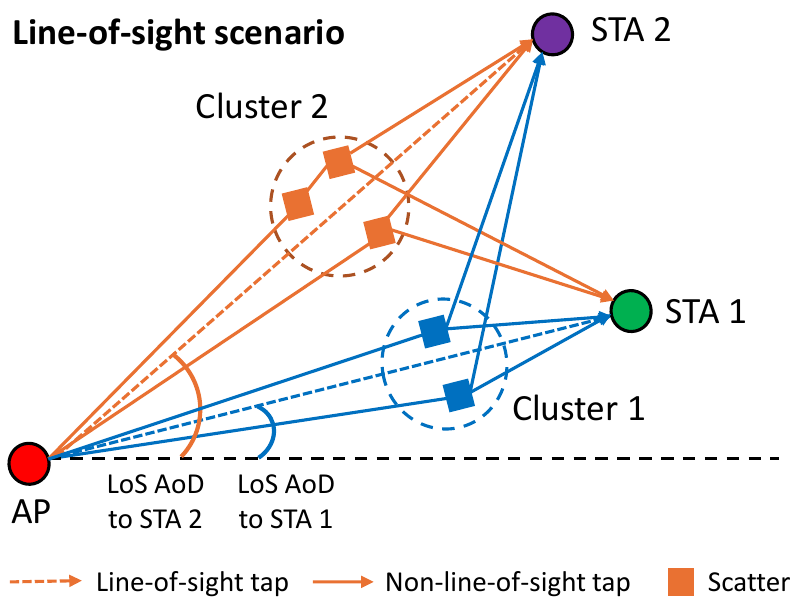}
    \caption{Modified IEEE 802.11ax Indoor Channel Model: DL SU (STA 1) and MU (STA 1 + 2) in Line-of-sight Scenario.}
    \label{fig:channelModel}
\end{figure}

We use the class of {\em cluster-based multipath fading channels} to model the practical environmental conditions for indoor Wi-Fi downlink operation. Such models were introduced by Saleh and Valenzuela, and extended/elaborated upon by many other researchers \cite{rappaport2024wireless}. In particular, IEEE 802.11ax indoor channel model \cite{liu2014ieee} is a typical cluster-based channel model that we have adapted by incorporating a parameter for user separation, as shown in Fig. \ref{fig:channelModel}. IEEE 802.11ax indoor channel model represents the propagation environment as a collection of scatterers grouped into clusters, where each cluster represents objects in the vicinity that act as a forward scattering source of rays that reach the receiver. Such clusters are typically represented via spatial-temporal models that capture the spatial characteristics of the environment, such as the transmit/receive antenna correlation and the distribution of objects, etc.  

A particular impact on our results arises from distinction between Line-of-sight (LoS)  and Non-line-of-sight (NLoS) scenarios as defined by 11ax channel model specification, depending on the relationship between the breakpoint distance \footnote{The breakpoint distance is defined as the distance that separates LoS and NLoS scenarios by characterizing different path loss exponents. } and the distance between AP and STA(s) \cite{liu2014ieee}:
\begin{itemize}
    \item {\em LoS scenario} (Fig. \ref{fig:channelModel}) occurs if the distance between AP and STAs is smaller than the breakpoint distance. The received signal at each STA include a LoS component and multiple multipath-induced NLoS components within a tapped delay-line model. This results in Rician fading multipath models where the first tap (corresponding to earliest arrival at each STA) is the LoS component. Therefore, the CSI obtained at each STA in such cases includes both LoS component and NLoS components with spatial characteristics \cite{liu2014ieee}; LoS CSI component depends on the transmit/receive steering vector parameterized by LoS angle of departure (AoD)/angle of arrival (AoA). Each NLoS CSI component depends on transmit/receive antenna correlation parameterized by NLoS mean AoD/AoA along with angular spread, and the spatial distribution of random scatterers within the cluster. The mathematical expression for LoS/NLoS CSI components can be found at \cite{schumacher2004description}. Since the first LoS tap signal is typically significantly stronger than NLoS signals, the LoS CSI component dominates the CSI obtained at each STA.  
    \item {\em NLoS scenario} occurs if the distance between AP and STAs is greater than the breakpoint distance; then the LoS tap signal at each STA in Fig. \ref{fig:channelModel} is blocked. Thus, the received signals at each STA are all NLoS (hence Rayleigh fading) and  the first NLoS tap signal's power is close to that of the other NLoS taps.      
\end{itemize}

\begin{figure}[t]
    \centering
\includegraphics[width=.38\textwidth]{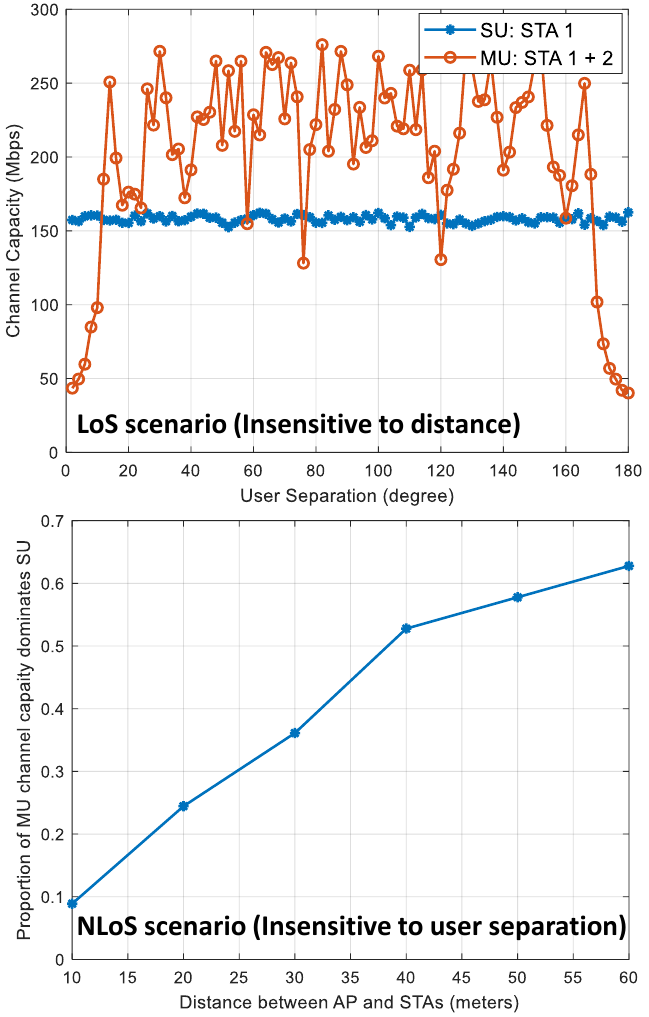}
    \caption{Channel Capacity impacted by Spatial Correlation. 20 dBm Transmit Power, 20 MHz Bandwidth, -174 dBm/Hz Noise Power Spectrum Density.}
    \label{fig:SUvsMU}
\end{figure}

% LoS $AOD_1 = 0^o$, LoS $AOD_2 \in [0^o, 180^o]$, $d_{BP} = 10$ m

\subsection{Spatial Correlation}
 Fig. \ref{fig:channelModel} includes an 8 x 8 uniform linear array (ULA)-based AP whose ULA antenna spacing is half wavelength as well as two 1 x 1 STAs. The sake of using a 2-user example here is to provide key insights for readers. Extending to a larger user number will be discussed in the next section. Thus AP transmits to STA 1 if MU-MIMO is turned off and to both STA 1 and STA 2 if MU-MIMO is turned on. The spatial geometry of STAs is characterized by their angle of departure (AoD), i.e., LoS AoD to STA 1 and LoS AoD to STA 2. The user separation between STA 1 and 2 is defined as the difference between LoS AoD to STA 2 and STA 1, respectively. To investigate the impact of user separation, we fix the angular geometry of cluster 1\footnote{Cluster 1's NLoS mean AoD equals the LoS AoD to STA 1; Cluster 2's NLoS mean AoD equals the LoS AoD to STA 2.} and STA 1, i.e., LoS AoD to STA 1 is set to $ 0^o$, and the LoS AoD to STA 2 varied between $0^o$ and $180^o$, thus the user separation between STA 1 and 2 ranges between $0^o$ and $180^o$.  

% \textbf{In the LoS scenario, the distance between AP and STAs ($d_{AP,STA}$) is usually small. The spatial correlation in the LoS scenario is not very sensitive to $d_{AP,STA}$ which can be negligible. Then user separation is the dominant feature that we will explore about the spatial correlation in the LoS scenario. In the NLoS scenario, the obtained CSI at STA is only contributed by NLoS taps' components that }  

\subsubsection{Dominant feature in the LoS scenario - User separation}
 Due to the small breakpoint distance, e.g., 10 meters, spatial correlation in the LoS scenario is not sensitive to the distance variation. Thus user separation is the single dominant feature that we explore in the LoS scenario, which is shown in Fig. \ref{fig:SUvsMU}. Consider a set of LoS scenarios where the distance is 8 meters and the granularity of user separation is $2^o$,  resulting in a total of 90 user separation scenarios. DL SU channel capacity dominates MU in 14\% scenarios of which 12\% scenarios lie in $0-12^o$ and $168-180^o$ user separation regions. Note that DL MU-MIMO channel capacity over $0-180^o$ user separation exhibits a symmetric channel capacity pattern, that can be attributed to the ULA characteristics where the LoS transmit/receive steering vectors of two STAs are identical at $0^o$ or $180^o$ user separation. Then, their dominant LoS CSI component determined by LoS transmit/receive steering vectors are also close for user separation regions close to $0^o$ or $180^o$. As a result, the corresponding V-matrices become highly correlated, incurring significantly higher inter-user interference than other user separation regions.  

% Thus, DL MU-MIMO operation in the LoS scenario benefits from the spatial diversity where a proper physical user separation between the clients is necessary \cite{cambium}.  However, most modern enterprise deployments of Wi-Fi involve a high density of users, and the resulting small user separation (smaller than $12^o$) between each device may not prevent high inter-user interference. This is why DL MU-MIMO usually suffers in dense networks.

\begin{figure*}[t]
    \centering
\includegraphics[width=.95\textwidth]{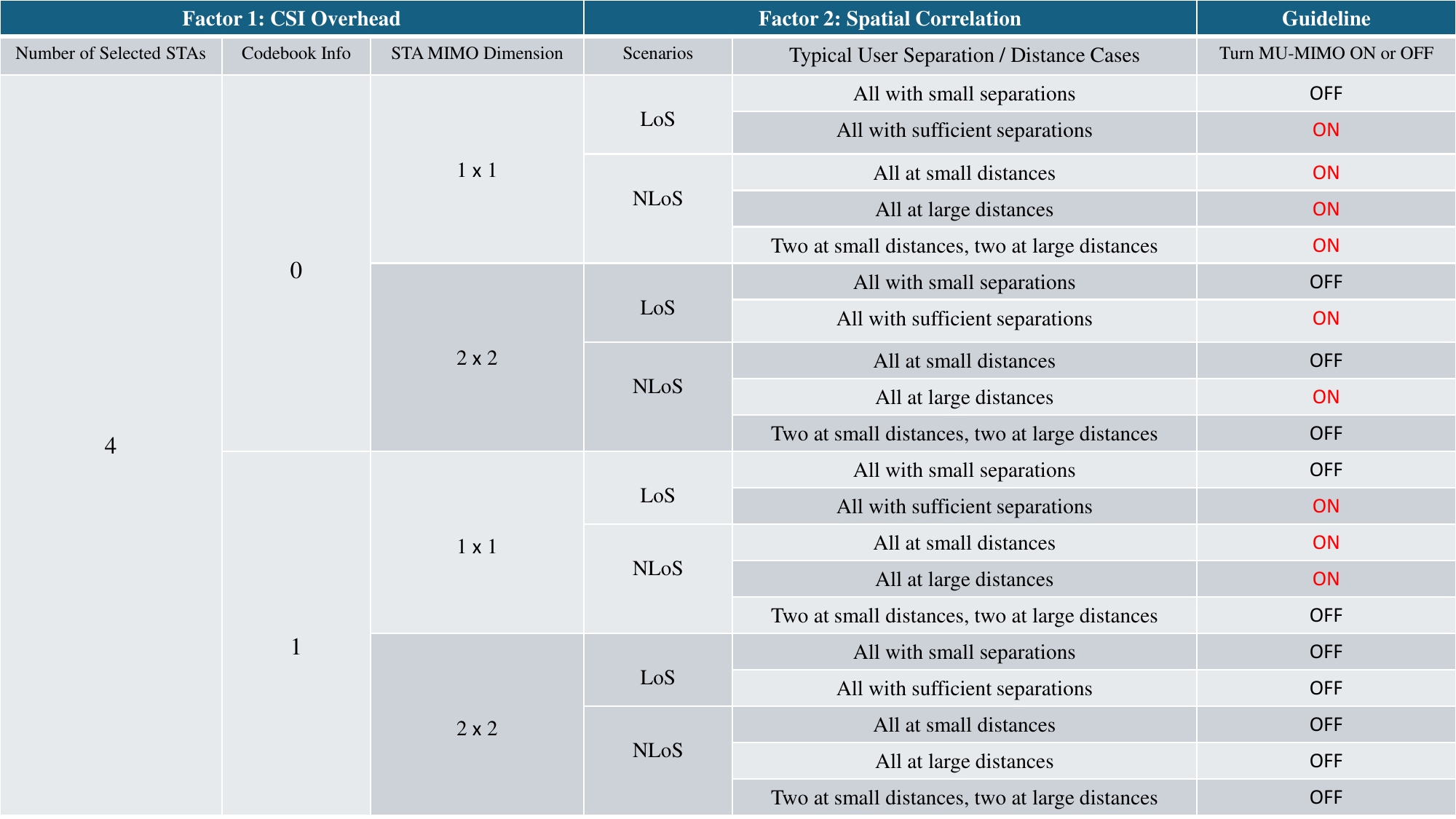}
    \caption{A 4-user Guideline Table for 8 x 8 AP under Modified IEEE 802.11ax Channel Model.  }
    \label{fig:table}
\end{figure*}

% In the $0-10^o$ user separation regions, the two STAs' LoS AoD/AoA difference is subtle, their dominant LoS CSI components become similar; While in the $170-180^o$ separation regions, as transmit/receive steering vectors are in $180^o$ periodicity of user separation, their dominant LoS CSI determined by transmit/receive steering vectors become also similar in the user separation close to $180^o$.
% Such characteristics lead to a symmetric back lobe to the main lobe, indicating that two STAs' beam patterns within $0-10^o$ user separation can be very similar to two STAs' beam patterns within $170-180^o$. 

\subsubsection{Dominant feature in the NLoS scenario - Distance between AP and STAs}
In the NLoS scenario, the obtained CSI includes only NLoS components, and each NLoS component (corresponding to a NLoS tap) is determined by the transmit/receive antenna correlation as well as the characteristics of scatterers. Since the latter such as their distributions, shapes and properties of materials are random, each NLoS tap consists of superposition of multiple independent individual path components leading to the complex Gaussian assumption  \cite{liu2014ieee}.  As a result, the inter-user/inter-steam interference can vary significantly as a function of STA distance in such cases (and is insensitive to user angular separation).

The spatial correlation as a function of distance in NLoS is shown Fig. \ref{fig:SUvsMU} for scenarios where the granularity of distance is 10 meters; the maximum distance for DL MU-MIMO operation with sufficiently high SNR at the STA is 60 meters. For each distance, 60 equally spaced user separations is used to calculate the proportion of scenarios in which MU channel capacity dominates SU. As shown, the proportion increases when distance increases, indicating that DL MU-MIMO benefits more than SU-MIMO at larger distance. In particular, MU becomes dominant over 50\% scenarios for distances greater than 38 meters. The larger the distance is, the more the multiple scattering, reflection, and diffraction paths that decorrelate the signals received by different users. Hence, the inter-user interference is effectively reduced with the increasing distance.

%This is because the larger the distance is, channel acts more like NLoS and hence the inter-user interference for MU-MIMO will become lower, which is the main reason that MU-MIMO has higher average channel capacity than SU-MIMO when the AP-STA distance is large even though user separation can be close to each other in one of many events.

% This is because the channel gain becomes lower with the increasing $d_{AP,STA}$ due to the pathloss. As the inter-user interference strength is scaled by the channel gain, the channel gain at large $d_{AP,STA}$ results in lower inter-user interference in DL MU-MIMO.       
\section{Design Guideline for DL MU-MIMO}
\label{secIV}
This section provides practical design guidelines that unify the underlying factors discussed in Section \ref{secII} and \ref{secIII}. For the same setup as Fig. \ref{fig:SUvsMU} used to obtain the channel capacity is now modified to derive the {\em effective} channel capacity as in Fig. \ref{fig:ohExample}. Meanwhile, We extend to a 4-user MU-MIMO operation, i,e, the user sub-set selection size is 4, indicating that upto 4 out of $X \geq 4$ STAs are selected if MU-MIMO is turned on. As the 4-user spatial correlation (where each is characterized by LoS/NLoS, user separation, and distance) results in a large set of scenario combinations, we thereby provide some typical scenarios due to the page limit. It should also be noted that real indoor channels might differ from the used channel model, that is, the exact spatial correlation threshold, such as $12^o$ user separation and $38$ meter distance in Section \ref{secIII}, used for turning on/off MU-MIMO might be different. However, real channels should have the same guideline trend as the used channel model under each operational scenario (without specifying specific thresholds) defined in Fig. \ref{fig:table}.  All results were implemented in Matlab using indoor MIMO WLAN channel models created by Schumacher et al, \cite{schumacher2004description}. 

As the main features regarding CSI overhead are codebook information for BF compression (i.e., codebook info 0 and 1) and STA MIMO dimensions (i.e., 1 x 1 and 2 x 2 STA), there are a total of 4 operational scenarios regarding CSI overhead. Meanwhile, we provide 5 typical operational scenarios (i.e., 2 LoS and 3 NLoS scenarios \footnote{For the operational scenario of two at small distances and two at large distances, AP is assumed to serve one of the STAs at small distances if MU-MIMO is turned off.}) regarding spatial correlation.  As a result, we provide guidelines for 20 scenarios unifying both CSI overhead and spatial correlation, as shown in Fig. \ref{fig:table}. Our conclusion for the 2-user case is that among these 20 scenarios, DL MU-MIMO can be turned on in 9 (45\%). According to the guideline table, \textbf{DL MU-MIMO can be turned on in the following scenarios:
\begin{itemize}
    \item 1 x 1 STAs with sufficient user separation in LoS;
    \item 2 x 2 STAs with codebook info 0 and sufficient user separation in LoS;
    \item 1 x 1 STAs in NLoS;
    \item 2 x 2 STAs with codebook info 0 and large distances in NLoS.
\end{itemize}}
Otherwise, DL MU-MIMO is suggested to be turned off, i.e., switch to DL SU-MIMO. Note that the condition for turning on DL MU-MIMO is more stringent for the 2 x 2 STA case, compared to the 1 x 1 STA case. This is because each spatial stream in the 2 x 2 STA case suffers more from interfering streams (self-interference from another stream for the same STA and/or streams from another STA) than the 1 x 1 STA case (only one interfering stream from another STA). Thus compared to the 1 x 1 STA case, MU-MIMO effective channel capacity is less likely to exceed SU-MIMO in the 2 x 2 STA case.

% Moreover, based on our study, the guideline table for 8 x 8 AP in Fig. \ref{fig:table} generally supports the 4 x 4 AP case whose guideline user separation/distance regions vary small compared to the 8 x 8 AP case.  

% \begin{table}[htbp]
%     \centering
%     \caption{Your table caption here}
%     \label{tab:mytable}
%     \begin{tabular}{|c|c|c|c|}
%     \hline
%     \textbf{Codebook} & \textbf{MIMO STA} & \textbf{STA Nums} & \textbf{SU/MU-MIMO Threshold } \\
%     \hline
%     \multirow{2}{*}{0} & \multirow{2}{*}{1x1} & 1 & Threshold for 1 \\
%     & & 2 & Threshold for 2 \\
%     & & 4 & Threshold for 4 \\
%     & & 8 & Threshold for 8 \\
%     \hline
%     \multirow{2}{*}{1} & \multirow{2}{*}{2x2} & 1 & Threshold for 1 \\
%     & & 2 & Threshold for 2 \\
%     & & 4 & Threshold for 4 \\
%     & & 8 & Threshold for 8 \\
%     \hline
%     \end{tabular}
% \end{table}

\section{Conclusion}
\label{secV}
This article provides new insights about the key underlying factors (i.e., CSI overhead and spatial correlation) that have resulted in AP vendors turning off the DL MU-MIMO feature as the default setting in their products. Based on our study and analysis,  guidelines as a function of operational scenarios is provided to address the fundamental question ``when DL MU-MIMO should be turned on/off" for current/next-generation Wi-Fi systems.
% This article investigates the DL SU/MU-MIMO throughput performance impacted by two factors, channel sounding overhead and spatial correlation of IEEE channel model with inter-user and inter-stream interference. A fundamental operation guideline for DL SU/MU-MIM is summarized in the end of the paper for WiFi system design.

% \appendix

% \input{appendix}

\bibliographystyle{IEEEtran}
\bibliography{reference}

\begin{IEEEbiographynophoto}
{Liu Cao} received the B.E. degree in electrical engineering from Jinan University, Guangzhou, China, in 2017, and the M.S. degree in electrical engineering from Northwestern University, Evanston, IL, USA, in 2019. He is currently working toward the Ph.D. degree in electrical \& computer engineering at the University of Washington, Seattle, WA, USA. His research interests include resource allocation in 5G NR Sidelink and Wi-Fi networks.
\end{IEEEbiographynophoto}

\begin{IEEEbiographynophoto}
{Lyutianyang Zhang} received the B.E. degree in electronic communication from The Australian National University and the Beijing Institute of Technology in 2017 and the M.S. degree and Ph.D. Degree in electrical and computer engineering at the University of Washington in 2019 and 2023. His research interests include resource allocation problems, PHY-MAC cross-layer algorithm design, and simulator design for Wi-Fi, cellular systems, and mobile edge computing networks. 
\end{IEEEbiographynophoto}

\begin{IEEEbiographynophoto}
{Sumit Roy} received the B.Tech. degree in electrical engineering from the Indian Institute of Technology Kanpur in 1983, the M.S. and Ph.D. degrees in electrical engineering from the University of California, Santa Barbara Santa Barbara, CA, USA, in 1985 and 1988, respectively, and the M.A. degree in statistics and applied probability in 1988. He is currently a Professor with the Department of Electrical and Computer Engineering, University of Washington. His research interests include fundamental design and evaluation of wireless communication and sensor network systems spanning a diversity of technologies and application areas: next-gen (5G and beyond) wireless LANs and cellular networks, heterogeneous network coexistence, spectrum sharing, software-defined radio platforms, and vehicular and sensor networking. He was elevated as a fellow of IEEE by the Communications Society in 2007 for contributions to multi-user communications theory and cross-layer design of wireless networking standards. 
\end{IEEEbiographynophoto}

\begin{IEEEbiographynophoto}
{Sian Jin} received the B.E. degree in electronic information engineering from University of Electronic Science and Technology of China in 2016 and Ph.D. Degree in electrical and computer engineering at University of Washington in 2022. During 2022 spring and summer, he worked at Princeton University as a postdoc. Since 2022 fall, he has been working at MathWorks, developing state-of-the-art signal processing algorithms.
His research interests include array processing, statistical signal processing, wireless localization, navigation and wireless communication.
\end{IEEEbiographynophoto}

\end{document}